\documentclass[useAMS,usenatbib]{mn2e}
\usepackage{graphicx}
\usepackage{amsmath,amssymb}
\usepackage{color,ulem}

\newcommand{\f}{\frac}

\newcommand{\BF}{\begin{figure}\begin{center}}
\newcommand{\EF}{\end{center}\end{figure}}
\newcommand{\BE}{\begin{equation}}
\newcommand{\EE}{\end{equation}}
\newcommand{\BEA}{\begin{eqnarray}}
\newcommand{\EEA}{\end{eqnarray}}

\newcommand{\IG}{\includegraphics}

\newcommand{\ms}{\textrm{M}_{\odot}}
\newcommand{\kms}{\textrm{km s}^{-1}}

\begin{document}

\title{Direct Gravitational Imaging of Intermediate Mass Black Holes in
Extragalactic Halos }

\author[Kaiki Taro Inoue, Valery Rashkov, Joseph Silk, Piero Madau]
{Kaiki Taro Inoue$^{1}$\thanks{E-mail:kinoue@phys.kindai.ac.jp}
Valery Rashkov$^2$\thanks{E-mail:valery@ucolick.org}
Joseph Silk $^{3,4,5}$\thanks{E-mail:silk@astro.ox.ac.uk}
and Piero Madau$^2$\thanks{E-mail:pmadau@ucolick.org}
\\
$^{1}$Department of Science and Engineering, 
Kinki University, Higashi-Osaka, 577-8502, Japan 
\\
$^{2}$Department of Astronomy and Astrophysics, University of California, Santa Cruz,
1156 High Street, Santa Cruz, CA 95064
\\
$^{3}$Institut d'Astrophsique de Paris, UMR 7095, CNRS, UPMC Univ. Paris VI, 98 bis boulevard Arago, 75014 Paris, France
\\
$^{4}$Department of Physics and Astronomy, The Johns Hopkins University Homewood Campus, Baltimore, MD 21218, USA
\\
$^{5}$Beecroft Institute for Particle Astrophysics and Cosmology, University of Oxford, Denys Wilkinson
Building, Keble Road, Oxford, OX1 3RH, UK}
\date{\today}
\pagerange{\pageref{firstpage}--\pageref{lastpage}} \pubyear{0000}
\maketitle
\label{firstpage}
\begin{abstract}
A galaxy halo may contain a large number of intermediate mass black holes (IMBHs)  
with masses in the range of $10^{2-6}\,\ms$.
We propose to directly detect these IMBHs
by observing multiply imaged QSO-galaxy or galaxy-galaxy strong
lens systems in the submillimeter bands with high angular resolution.
The silhouette of an IMBH in the lensing galaxy halo would appear as either 
a monopole-like or a dipole-like variation at the scale of the 
Einstein radius against the Einstein ring of the dust-emitting region
surrounding the QSO. We use a particle tagging
technique to dynamically populate a Milky Way-sized dark matter halo with 
black holes, and show that the surface mass density and number
density of IMBHs have power-law dependences on the
distance from the center of the host halo if smoothed on a scale 
of $\sim 1\, \textrm{kpc}$. Most of the black holes orbiting close to the 
center are freely roaming as they have lost their dark matter hosts 
during infall due to tidal stripping. Next generation submillimeter telescopes with 
high angular resolution ($\lesssim 0.3\,$mas) will be capable of 
directly mapping such off-nuclear freely roaming 
IMBHs with a mass of $\sim 10^{6}\,\ms $ in a lensing galaxy
that harbours a $O(10^{9})\,\ms $ supermassive black hole in its nucleus.
\end{abstract}

\section{Introduction}

Recent observations of off-nuclear ultraluminous X-ray sources (ULXs) suggest the presence of 
intermediate mass black holes (IMBHs) not only in the neighborhood of the galaxy 
nucleus but also in star clusters far out in the galactic halo 
\citep{matsumoto2001, roberts2004, farrell2009, jonker2010}. A large population of IMBHs 
might reside inside a galaxy halo, perhaps the leftover population of initial ``seed" holes that 
never grew into the supermassive variety (SMBHs\footnote{Here a ``SMBH"
means a black hole (BH) residing at the center of a main parent halo. If the
host halo habouring the BH  belongs to a more massive parent halo,
we call it an ``IMBH''. }) hosted today in the nuclei of massive galaxies. 
The mechanism of seed formation is unknown. Seed holes may be produced by the direct collapse 
of $10^4-10^6\,\ms$ primordial gas clouds, by the collapse of the first nuclear star clusters,  
or be the remnants of  $10^2\,\ms$ Population III stars \citep[e.g.][]{loeb94,madau01,devecchi2009}.
If we could directly observe the abundance, spatial distribution, and masses of IMBHs 
inside extragalactic halos, we would be able to constrain the process of 
SMBH seed formation that has hitherto been veiled in mystery. 

We propose here to directly detect IMBHs
by observing multiply imaged QSO-galaxy or galaxy-galaxy
lens systems in the submillimeter 
band with high angular resolution ($\lesssim 0.3\,$mas).
Provided that the lensed source is sufficiently extended and 
the IMBHs inside the lensing galaxy
are massive enough, then their gravitational force would significantly distort the 
lensed images \citep{inoue2003,inoue2006}. 
Moreover, any star cluster or dark matter subhalo that surrounds an IMBH would
also distort the lensed images at angular scales larger than the Einstein radius of the IMBHs.
Using local distortions in the surface brightness of the lensed images, 
one can directly measure the position, mass scale, and gravitational potential surrounding 
the IMBHs.

In this paper, we estimate the feasibility of mapping IMBHs in forthcoming observations. In Section 2, we 
describe a particle tagging technique to dynamically populate the N-body {\it Via Lactea II} (VLII) 
extreme-resolution simulation with IMBHs, and derive semi-analytic formulae for describing the surface mass and number density 
of IMBHs in the host halo. In Section 3, we estimate 
the strong lensing probability due to IMBHs in a lensing galaxy. We then describe the extended source effects and observational
feasibility of direct detection in the QSO-galaxy lensing system RXJ1131-1231.
We summarise our results in Section 4.
In what follows, we assume a concordant 
cosmology with a matter density $\Omega_m=0.272$, a baryon density 
$\Omega_b=0.046$, a cosmological constant $\Omega_\Lambda=0.728$,
and a Hubble constant $H_0=70, \textrm{km}/\textrm{s}/\textrm{Mpc}$, 
which are obtained from the observed 
CMB, the baryon acoustic oscillations, and measurement of $H_0$.   

\section{A population of IMBHs}

\subsection{Simulation and tagging technique}

VLII, one of the highest-resolution N-body simulations of the assembly
of a Milky Way-sized galaxy halo to date \citep{diemand2008}, provides an ideal set-up for the
study of formation of IMBHs. VLII follows the hierarchical assembly of a dark halo of mass
$M_{200} = 2 \times 10^{12}\ms$ at redshift $z=0$ within $r_{200}$ = 402\,kpc (the radius which encloses an
average mass density 200 times the mean cosmological matter density), with just over a billion particles and a force resolution of 40 pc. 
The simulation was performed with the parallel tree-code PKDGRAV2 \citep{stadel01}. PKDGRAV2 uses a fast multipole expansion technique 
in order to calculate the forces with hexadecapole precision, and an adaptive leapfrog integrator. Expected to harbor a thin disk galaxy, VLII was
selected to have no major mergers after $z = 1$. At the present epoch, VLII has a cuspy density profile and exhibits rich galactic
substructure - the main host's halo contains over 20,000 surviving subhalos with masses greater than $10^6\ms$.

Central black holes are added to subhalos following the
particle tagging technique detailed in \citet{ras13} and quickly summarized here. In each of 27 snapshots of the simulation,
choosen to span the assembly history of the host between redshift $z=27.54$ and the present, all subhalos are identified and linked
from snapshot to snapshot to their most massive progenitor: the subhalo tracks built in this way contain all the time-dependent structural
information necessary for our study. We then identify the simulation snapshot in which each subhalo reaches its maximum mass $M_{\rm halo}$
before being accreted by the main host and tidally stripped. In each subhalo, the 1\% most tightly bound dark matter
particles are then ``tagged" as stars at infall, following a set of
simple prescriptions calibrated to reproduce the observed luminosity function of Milky Way satellites and the concentration of their stellar
populations \citep{ras12}. We then measure the stellar line-of-sight velocity dispersion, $\sigma_*$, in each subhalo, and tag {the most tightly bound
central particle} as a black hole of mass $M_{\rm BH}$ according to an extrapolation of the $M_{\rm BH}-\sigma_*$ relation of \citet{tremaine2002},
\begin{equation}
M_{\rm BH} = 8.1 \times 10^6\,\ms \left(\frac{\sigma_*}{100\,\kms}\right)^4.
\label{msigma}
\end{equation}
For stellar systems with $\sigma_*\le 6\,\kms$, a minimum seed hole mass of $100\,\ms$ is assumed.
Any evolution of the tagged holes after infall is purely kinematical in character, as their satellite hosts are accreted and disrupted in an
evolving Milky Way-sized halo. Black holes do not increase in mass after tagging, and are tracked down to the $z=0$ snapshot. We assume 
that IMBHs only form in subhalos with a mass at infall $>M^s_{\rm min}=10^7\ms$.

The merging process that produces each subhalo prior to its infall into the main host halo is important, as black hole binaries may form in 
the process. The asymmetric emission of gravitational waves produced during the coalescence of a binary black hole system is known to impart 
a velocity kick to the system that can displace the hole from the center of its host. The magnitude of the recoil will depend on the binary mass ratio and 
the direction and magnitude of their spins, and follows the prescriptions detailed in \citet{guedes2011}. When the kick velocity is larger 
than the escape speed of the host halo, a hole may be ejected into intergalactic space before becoming a Galactic IMBH. 
IMBHs that would have formed in halos whose merger history points to a kick event are therefore excluded from the final catalog (see \citealt{ras13} 
for details). 

Since some of the subhalos will fall on trajectories that bring them closer to the center of the main VLII host, they might
be completely disrupted after infall, leaving  freely-roaming (``naked") holes. Depending on the minimum halo mass that allows the formation of an IMBH, 
the numbers of formed, kicked, and stripped IMBHs will be different. Figure 1 shows an image of the projected distribution of IMBHs in today's Galactic 
halo according to the above prescriptions. We find 1,070 naked IMBHs and 1,670 holes residing in dark matter satellites that survived tidal
stripping. There are 50 IMBHs more massive than $10^3\,\ms$, and about 2,500 holes at the minimum mass of $100\,\ms$ (of which 950 are naked).  
Naked holes are more concentrated towards the inner halo regions as a consequence of the tidal disruption of infalling satellites. 
Indeed, within 10 kpc, most MBHs are naked.

\begin{figure*}
\begin{center}
\IG[width=18cm]{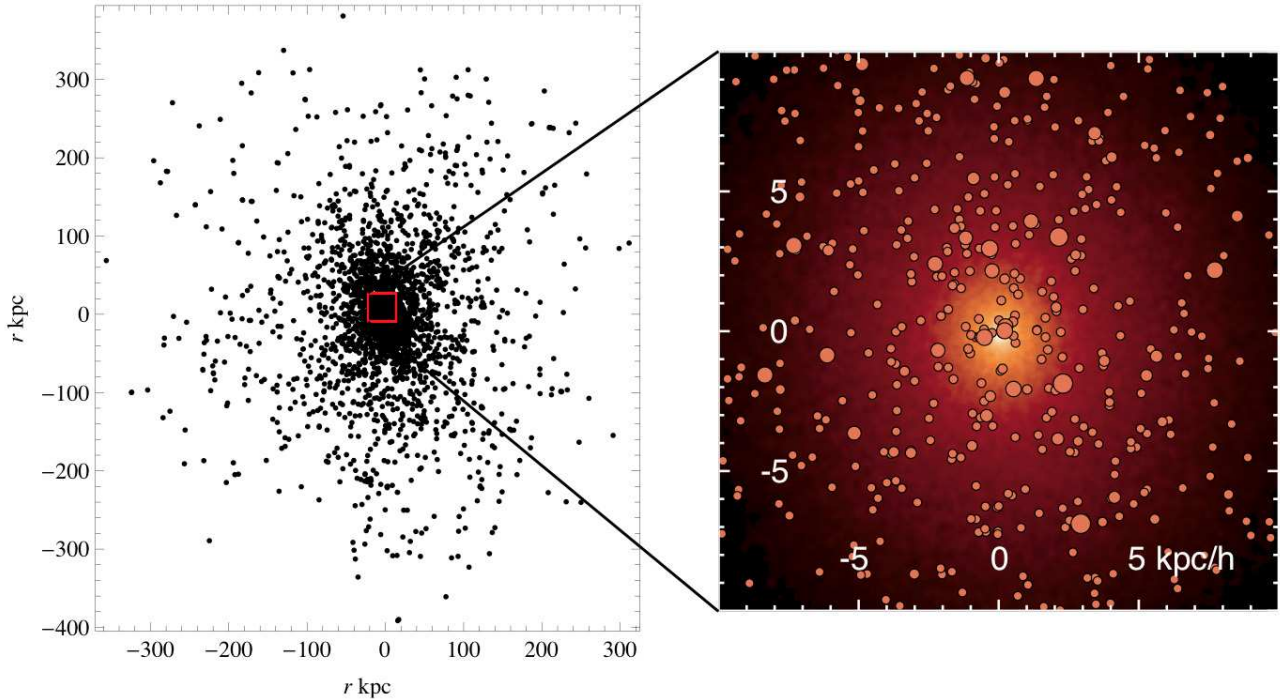}
\caption{Plot of the distribution of IMBHs in the simulated Milky Way-sized halo. Each dot represents 
the position of an IMBH (left). The 
threshold of the minimum subhalo mass 
required for the formation of a seed hole is 
assumed to be $M^s_{\min}=10^7\ms$. The projected density of dark matter in
the VLII simulation is plotted in a zoom-up image, where the size 
of each disk is proportional to the $\log_{10}$ of the BH mass (right). 
 }
\end{center}
\end{figure*}

\subsection{Semi-analytic modeling}

To generalize the numerical results of the previous section, we shall assume in the 
following that the distribution of IMBHs
inside a lensing galaxy halo is spherically symmetric,
and that the surface mass density $\sigma_m$ and 
the surface number density $\sigma_n$ are described by
the following ``universal profiles'',
\BEA
\sigma_m(r) &=&\frac{ \sigma_m(0)}
{((r/r_c)^2+1)\exp[r/r_*]},
\label{eq:mass}
\\
\sigma_n(r)&=&\frac{\sigma_n(0)}{(r/r_c+1)\exp[r/r_*]},
\label{eq:surface}
\EEA
where $r_c$ represents a core radius and $r_*$ denotes 
a cut-off radius at which the density starts to decay exponentially
with increasing $r$. The ``universal profiles'' are much steeper than
those of a singular isothermal sphere (SIS). 
Subhalos hosting IMBHs can be massive 
enough to experience dynamical friction, spiral in toward the center of the 
main host, be totally stripped of their dark matter, and deposit a naked 
IMBH into the center of the main  host.
As shown in Figure 2, 
most of IMBHs at $r \lesssim 10\,\textrm{kpc}$ are naked. 
We find that the surface mass density at $r\lesssim 10\,\textrm{kpc}$ is not
significantly affected by $M^s_{\min}$. 

The constants $\sigma_m(0)$ and $\sigma_n(0)$ can 
be estimated as follows. First, each halo is approximated as a spherically symmetric 
object with virial radius $r_{200}$. We assume that within $r_{200}$ the halo density profile
is given by that of an SIS with one
dimensional velocity dispersion $\sigma_v$. Then we have
\BE
r_{200}(z)=\frac{\sigma_{v}}{H(z)\sqrt{50\, \Omega_m(z)}},
\EE 
where $\Omega_m(z)$ and $H(z)$ are the matter 
density parameter and the Hubble parameter at redshift $z$, respectively.
The initial size of the halo is then
\BE
r_{ini}\approx 200^{1/3} r_{200}.
\EE
The correlation between the stellar velocity dispersion
$\sigma_{v} $ and mass $M_{\textrm{SMBH}}$ of the supermassive black
hole at the center is approximately given by
\BE
\biggl(\frac{M_{\textrm{SMBH}}}{10^8\,\ms}\biggr) 
\sim \beta \times \biggl ( \frac{\sigma_{v}}{200 \, \textrm{km/s}} \biggr)^\alpha,
\label{eq:msigma}
\EE
where $\beta=O(1)$ and $4<\alpha<5$. 
The mass of the SMBH at the nucleus of the Milky Way
is $M_{\textrm{SMBH}}=4 \times 10^6\, \ms$. Then, the 
velocity dispersion of the spheroidal component 
is $\sigma_{v}=88\, \textrm{km/s}$ and 
$r_{200}=3.4\times 10^2\,\textrm{kpc}$ at $z=0$ provided
that $\alpha=4.24$ and $\beta=1.32$ \citep{gultekin2009}.
We also assume that the total mass of IMBHs
within $r_{200}$ is 
given by $M_{\textrm{IMBH}}(<r_{200})=f M_{\textrm{SMBH}}$, where $f=0.1-1$. 
Then from equations (\ref{eq:mass}) and  (\ref{eq:msigma}), we have
\BE
\sigma_m(0) \approx \frac{\beta f  
\bigl(\sigma_v/200\,\textrm{km/s}
 \bigr)^\alpha }{2 \pi r_c^2 \bigl(\ln{(r_{200}/r_c)}+1/2 \bigr)} \times
 10^8\,\ms.
\label{eq:sigma0}
\EE 
If we adopt $f=0.2$, the analyticaly estimated surface mass density
$\sigma_m$ fits the simulated values well (see Fig. 2).  
We find that the surface mass density in the neighborhood of the center 
does not change much even if one changes the threshold $M^s_{\textrm{min}}$.

Second, we assume that 
the seed of an IMBH is formed at a 
redshift of $z=20$ inside a host halo with a mass of 
$M>M^s_{\min}$, and each seed grows almost independently. 
Then the approximated virial radius
of the halo is $\tilde{r}_{200}(z=20)=3.3\times 10^2\, \textrm{pc}$.
Assuming an SIS profile, the maximal circular velocity
of a host halo is estimated as $V_{max}=\sqrt{2} \sigma_v=10H\tilde{r}_{200}=12\,\textrm{km/s}$. 
At $z=20$, the number density of subhalos with maximal velocity larger than
$V_{max}$ is approximately given by
\BEA
n(>V_{max}) &=&\frac{A}{V_{max}^3},
\nonumber
\\
A(z=20)&=&1.43\times 10^5 \biggl(h^{-1}\textrm{Mpc}/(\textrm{km/s})
\biggr)^{-3},
\EEA
provided that $V_{max} \ll 1000\, \textrm{km/s}$ \citep{klypin2011}.  

For the Milky Way-sized halo in VLII, the initial radius 
is estimated as $r_{ini}(z=20)=1.65\,\textrm{Mpc}/h$. Assuming that all the halos with mass $>10^7\, \ms$
contain an IMBH seed, the total number of IMBH seeds inside a
lensing galaxy halo is estimated as
\BEA
N(>V_{max}=12\,\textrm{km/s})&=&\frac{A(z=20)}{(V_{max}=12\,\textrm{km/s})^3}
\times
\frac{4}{3}\pi r_{ini}^3
\nonumber
\\
&=&1.8\times 10^3.
\EEA
Note that $N(>V_{max})$ is proportional to $r_{200}^3$. 
If the total number of IMBHs within $r_{200}$ 
does not change much, then the surface number density 
at the center is 
\BE
\sigma_n(0)\approx \frac{N(>V_{max})}{2 \pi r_{200}r_c}.
\label{eq:sigman}
\EE  
Thus $\sigma_n(0)$ is proportional to $r_{200}^2$ or
$\sigma_v^2$ if $r_c$ does not depend on $r_{200}$. 
As shown in Figure 3, the ``universal profile'' 
fits the simulation well if the total number of IMBHs
coincides with that of the simulation. If $\sigma_n(0)$ 
in equation (\ref{eq:sigman}) is used, however, the surface number density 
is systematically reduced by a factor of $2-5$ in comparison with 
the simulation. This difference may be due to the strong clustering of 
massive halos.  
\BF
\IG[width=8cm]{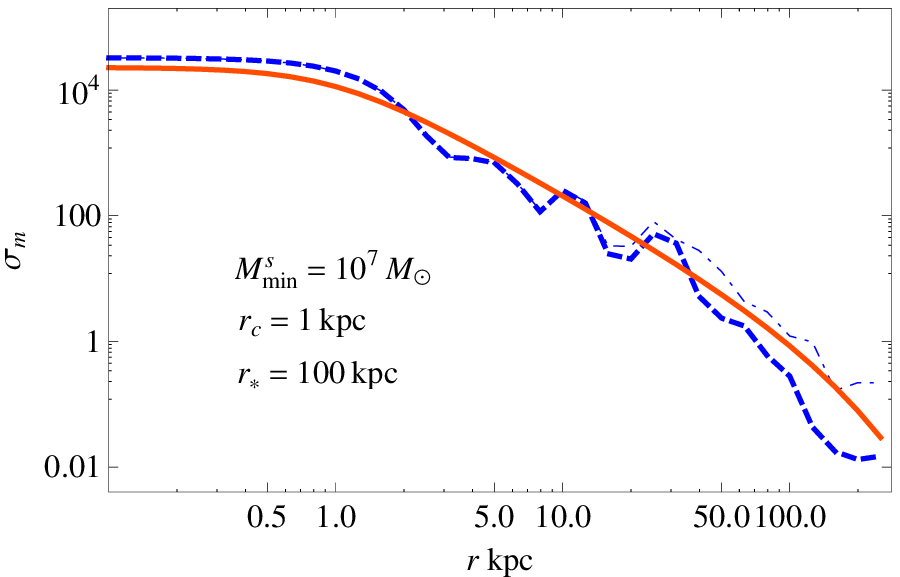}
\IG[width=8cm]{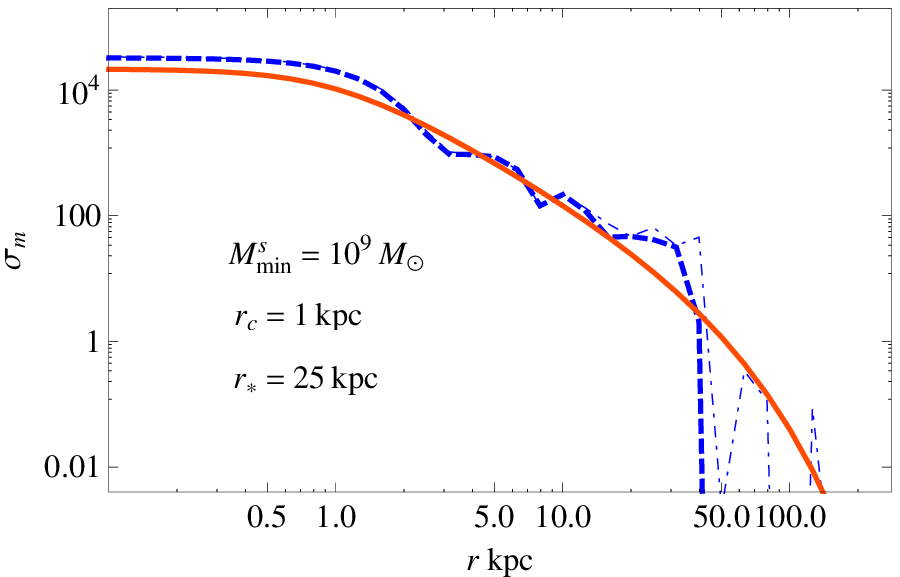}
\caption{Surface mass density of IMBHs in the simulated Milky Way-sized halo. 
The dashed and dot-dashed curves correspond to the naked and the total
 (naked plus hosted-in-substructure) IMBHs in our numerical simulations, respectively.
The solid curves represent fitted ``universal profiles'' in which the total mass of IMBHs 
coincides with that of our simulated results. This curve is also
obtained if we assume $f=0.2$.   
The distribution of simulated IMBHs is smoothed by a Gaussian window function 
$W(r)=\exp[-r^2/(2r_c^2)]$, where $r_c=1\,\textrm{kpc}$. 
The total masses of IMBHs (naked plus hosted-in-substructure)
are $9\times 10^5 \ms$ (top) and $8 \times 10^5\ms$ (bottom).
 }
\EF
\BF
\IG[width=8cm]{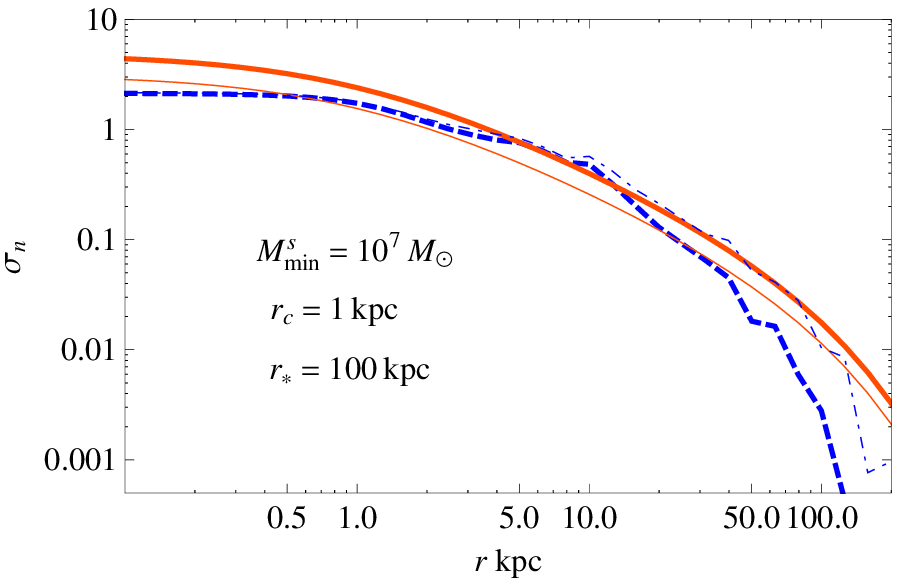}
\IG[width=8cm]{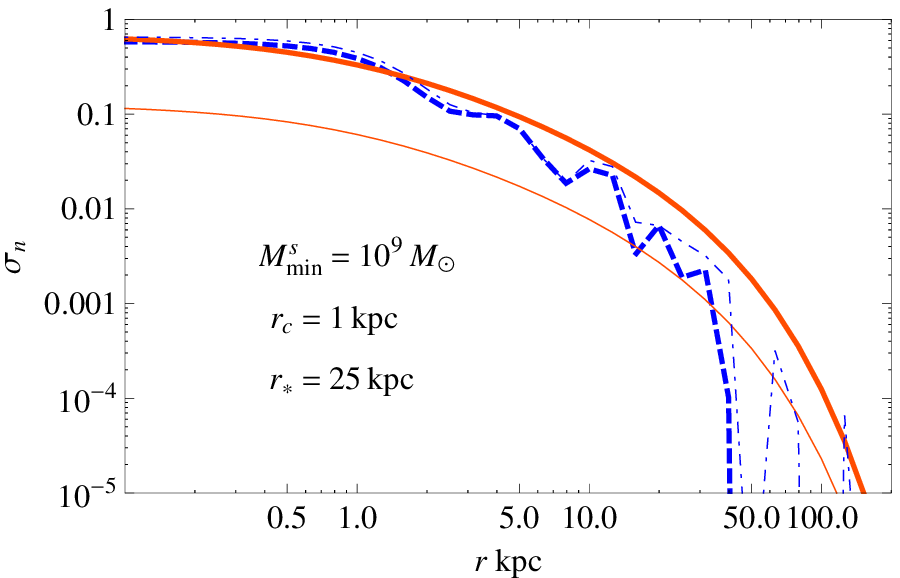}
\caption{Surface number density of IMBHs in the simulated Milky Way-sized halo.
The dashed and dot-dashed curves correspond to the naked and the
total(naked plus hosted-in-substructure) IMBHs in our numerical simulations, respectively. The solid
curves represent fitted ``universal profiles'' 
in which the total number of IMBHs coincides with  
that of our simulated results (bold solid) and that of our semi-analytically
estimated values (thin solid). 
The distribution of simulated IMBHs is smoothed by a Gaussian window function 
$W(r)=\exp[-r^2/(2r_c^2)]$, where $r_c=1\,\textrm{kpc}$. }
\EF
\section{Lensing effect}
\subsection{Lensing probability}
The Einstein angular radius $\theta_E$ of a point mass with a mass 
$M$ is written in terms of angular diameter distances: to the lens $D_L$,
to the source $D_S$, and between the lens and the source 
$D_{LS}$ as 
\BE
\theta_E\sim 3 \times  \Biggl ( \f{M}{10^6 M_{\odot}}
\Biggr)^{\f{1}{2}} \Biggl ( \f{D_L D_S /D_{LS}}{\textrm{Gpc}}
\Biggr)^{-\f{1}{2}} \textrm{mas}.
\EE
Therefore, a radio interferometer with resolution of 
3 mas can easily resolve the distortion of an image within 
the Einstein ring for a point mass $M \sim 10^6 \ms$. 

The strong lensing cross section due to an IMBH
is proportional to $M$. Therefore, the lensing probability
$p$ is given by the ratio between the surface mass density
of IMBHs and that of a lensing galactic halo at $r=r_E=D_L\theta_E$.
From equations (\ref{eq:mass}) and (\ref{eq:sigma0}), we have
\BE
p=\frac{\sigma_m(\textrm{IMBH})}{\sigma_m(\textrm{SIS})}
\biggl |_{r=r_E} \propto \frac{\sigma_v^\alpha}{\sigma_v^2 r_E}\propto f \sigma_v^{\alpha-4},
\EE
where $\sigma_m(\textrm{IMBH})$ and $\sigma_m(\textrm{SIS})$ denote
the surface mass density of IMBHs and that of an SIS, respectively.
Similarly, the mean mass $\bar{M}$ of an IMBH at $r=r_E$ satisfies  
\BE
\bar{M}=\frac{\sigma_m(\textrm{IMBH})}{\sigma_n(\textrm{IMBH})}
\biggl |_{r=r_E}
\propto f \sigma_v^{\alpha-4}.
\EE
Thus the lensing probability and the Einstein radius are larger 
for halos with larger velocity dispersion as long as $\alpha>4$
and equations (\ref{eq:mass}) and (\ref{eq:surface}) 
hold. As the Einstein radius of an IMBH is proportional to $D_L^{-1/2}$,
lens systems with small $D_L$ are more suitable as targets.
\subsection{Extended source effect}
If a perturber is spatially extended, then
the lensing effect is different from that of a point mass.
The density profile of a perturber can be reconstructed
from a local mapping between the observed image
and the non-perturbed image obtained from the 
prediction of the macrolens. In fact, the power of the 
radial density profile of the perturbers can be reconstructed from the 
perturbed images within the Einstein ring of the perturber
\citep{inoue2005b}. In this way, one could make a distinction
between an IMBH and a CDM subhalo. 
Furthermore, from distortion outside the Einstein ring
of the perturber, the degeneracy 
between the perturber mass and the distance 
can be broken provided that the Einstein radius
of the perturber is sufficiently smaller than that 
of the primary macrolens \citep{inoue2005a}.
The precise measurement of spatial variation in the surface
brightness of lensed images provides us with plenty of information about the 
mass, abundance, and spatial distribution of IMBHs.

\subsection{Simulation for RXJ1131-1231}
To estimate the observational feasibility of direct detection of IMBHs,  
we adopt a QSO-galaxy 
lensing system RXJ1131-1231 as a target 
since this system has a massive lensing halo at relatively
small redshift.
The redshifts of the source and the lens
are $z_S=0.658$ and $z_L=0.295$ \citep{sluse2003}, respectively. 
To model the macro-lens, we adopt
a singular isothermal ellipsoid (SIE) in a constant
external shear field in which the
isopotential curves in the projected surface perpendicular to the
line-of-sight are ellipses \citep{kormann1994, inoue2012}.
The IMBHs inside the lensing halo are modeled as point masses. 
Using the observed mid-infrared fluxes, the position of the lensed images
and the centroid of the lensing galaxy, the velocity dispersion 
for the best-fit model is estimated as $\sigma_v=3.5\times 10^2\,\textrm{km/s}$.
From the $M-\sigma$ relation with $\alpha=4.24 $ and $\beta=1.32$ \citep{gultekin2009}, the total mass of IMBHs within $r_{200}$ is $M_{\textrm{IMBH}}=1.5 f\times 10^9\,\ms$.
Then the lensing probability is $p=5f\times 10^{-4}$. The mean
mass of the IMBH at $r=r_E=5.6\,\textrm{kpc/h}$ is $\langle
M_{\textrm{IMBH}} \rangle=2f\times10^5\,\ms $
and the corresponding Einstein radius is $\theta_E=\sqrt{f}\,\textrm{mas}$. 
Therefore, if the angular resolution is $< \sqrt{f}\, \textrm{mas}$, 
we would be able
to detect an imprint of IMBHs in the lensed Einstein ring of
dust emission if the lensed image has an area $>2000 f^{-1}\, \textrm{mas}^2$. 

To estimate observational feasibility, we use the simulated data
of IMBHs for a Milky Way-sized halo with $M^s_{\min}=10^7\,\ms$ 
and scale up the mass of each IMBH
by a ratio between the total mass of the SMBH for RXJ1131-1231 
($=1.5\times 10^9\,\ms$) and
that for the Milky Way-sized halo ($=4\times 10^6\,\ms$). 
In this model, we find that the 
ratio between the mass fraction of all IMBHs to that of a SMBH 
in the center is $f=0.2$. The most massive IMBH has a mass of
$M_{\textrm{IMBH}}=7\times10^7\,\ms$. 
Taking into account the logarithmic correction to $\sigma_v$ in equation
(\ref{eq:sigma0}), the distance of each particle from the center
is scaled up by a factor 
\BEA
\gamma&\approx&\biggl(\f{\ln(\sigma_v(\textrm{RXJ1131})/(\sqrt{50}H(z=0.295)r_c))}{
\ln(\sigma_v(\textrm{MW})/(\sqrt{50}H(z=0)r_c)) } \biggr)^{1/2}
\nonumber
\\
&=& 1.05,
\EEA
where we assume that $r_c=1\,\textrm{kpc}$. As shown in Figure 4, we find that  
massive ``naked'' IMBHs with masses in the range of 
$O(10^{5-6})\,\ms$ are observable if the radius of the dust emitting region
around the QSO is $\sim 500\, \textrm{Mpc}$
and the angular resolution is $< 1\, \textrm{mas}$.
The area of the lensed arc is found to be 
$\sim 6\times 10^5 \textrm{mas}^2$. Thus we expect $>O(10)$ ``naked'' IMBHs
inside the lensed arc if $f>0.2$.

Even if IMBHs reside in places far from the brightest
part of the Einstein ring of dust emission, the surface brightness 
of the lensed 
image that is sufficiently close to the position of an IMBH is the same
as that of the brightest region in the Einstein ring.
It is easily distinguished from other sources as the spatial
gradient of the surface brightness is extremely large near the
position of an IMBH.

\begin{figure*}
\centering
\IG[width=16cm]{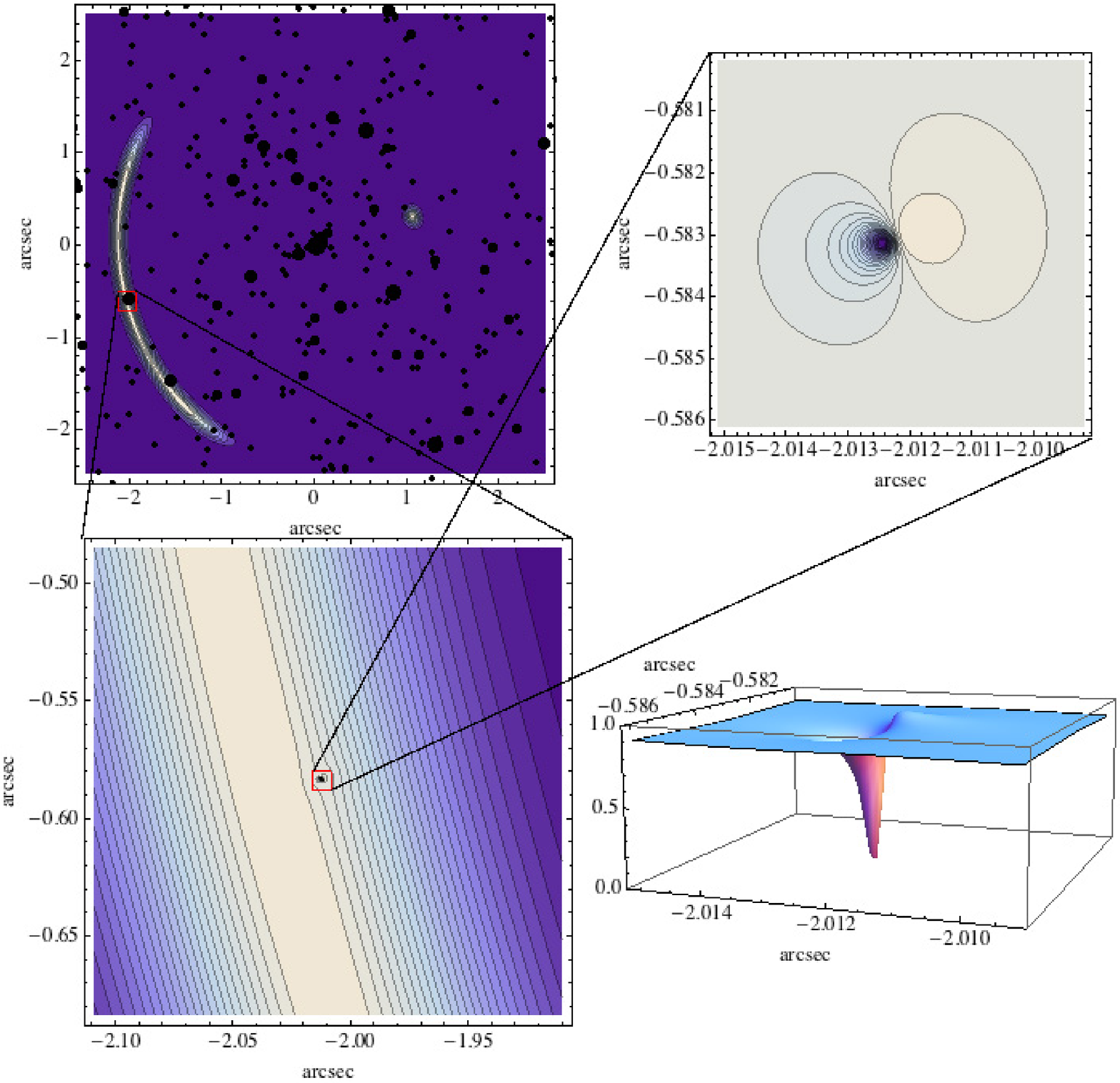}
\caption{Simulated lensed images
of RXJ1131-1231.
The black dots in the top left panel
show the positions of IMBHs in the lensing galaxy and
the dot size is proportional to $\log_{10}(M_{\textrm{IMBH}})$. 
The other panels 
show the contour and 3D maps of the  
surface brightness centered at the angular position of 
an IMBH with a mass of $1.4 \times 10^6\, \ms$ that produces
a ``black hole''.
We assume that the dust emitting region has a 
circular gaussian luminosity profile with a $1\,\sigma$ radius
$r=500\,$pc. The numbers of contours representing the 
surface brightness are 10 (top left), 25 (bottom left), 25 (top right).}
\end{figure*}

\section{Summary}
Based on numerical simulations of IMBH formation in a Milky-way sized halo,    
we have found that the surface mass density and number
density of IMBHs have power-law dependences on the
distance from the center of the host halo if smoothed on scales of 
$\sim 1\,\textrm{kpc}$. Most of the black holes orbiting close to the
center are freely roaming as they have lost their surrounding dark mass 
during infall due to tidal stripping. Assuming 
the surface mass density and number
density have such power-law dependences,  
the strong lensing probability due to 
free roaming IMBHs in a massive lensing galaxy 
such as RXJ1131-1231 is $\sim O(10^{-4})$. 
The next generation submillimeter telescopes with 
high angular resolution ($\lesssim 0.3\,$mas) will be  capable of 
directly mapping these freely roaming 
IMBHs with a mass of $\sim 10^{6}\,\ms $.

In addition to IMBHs in the lensing galaxy, SMBHs with a mass of 
$\gtrsim 10^6\,\ms $ inside halos 
in the line-of-sight may also be observable, especially in 
systems that show anomalies in the flux ratios \citep{inoue2012}.
In this case, some distortion in the lensed image due to the host halo 
is expected. 

By measuring the local distortion of lensed 
extended images with high angular resolution, 
we will be able to determine the mass, abundance,
and density profile of the IMBHs present in the lensing galaxy.
Direct detection of IMBHs will shed  new light on the 
formation process of SMBH seeds which 
has hitherto been shrouded from view.

\section{Acknowledgments}
This work was supported by 
the NSF through grant OIA-1124453, and by NASA through 
grant NNX12AF87G.
This research was also supported in part by ERC project  267117 (DARK) hosted by  Universit\'e Pierre et  Marie Curie - Paris 6.

\bibliographystyle{mn2e}

\bibliography{direct-image-IMBH}
\label{lastpage}

\end{document}